\documentclass[11pt]{article}

\usepackage{palatino}
\usepackage{amsmath}
\usepackage{amsfonts}
\usepackage{amssymb}
\usepackage{latexsym}
\usepackage{geometry}             
\usepackage{epstopdf}
\usepackage{graphicx}
\usepackage{hyperref}

\newcommand{\spa}[1]{\mathcal{#1}}
\newcommand{\Tr}{\mathrm{Tr}}
\newcommand{\half}{\frac{1}{2}}

\newcommand{\calA}{\mathcal{A}}
\newcommand{\calB}{\mathcal{B}}

\newcommand{\calM}{\mathcal{M}}
\newcommand{\calH}{\mathcal{H}}

\newcommand{\tr}{\mathrm{Tr}}
\newcommand{\ket}[1]{| #1 \rangle}
\newcommand{\bra}[1]{\langle #1 |}

\newcommand{\ketbra}[2]{\ket{#1} \bra{#2}}
\newcommand{\kb}[1]{\ketbra{#1}{#1}}
\newcommand{\braket}[2]{\langle #1 | #2 \rangle}

\newcommand{\inner}[2]{\langle #1, #2 \rangle}

\newcommand{\set}[1]{\left\{ #1 \right\}}

\newcommand{\Pos}{\mathrm{Pos}}

\newcommand{\R}{\mathbb{R}}
\newcommand{\Z}{\mathbb{Z}}

\newcommand{\norm}[1]{\left\| #1 \right\|}

\newcommand{\chews}[2]{{ #1 \choose #2 }}
\newcommand{\zo}{\{0,1\}}
\newcommand{\eps}{\varepsilon}

\newcommand{\knfOT}{ {n \choose k}\mbox{-}\mathrm{fOT}}

\newcommand{\fOT}[2]{{{#2} \choose {#1}}\mbox{-} \mathrm{fOT}}

\newcommand{\comment}[1]{}

\newcommand{\notabortOTA}{\lnot \bot^{OT}_A}
\newcommand{\notabortCFA}{\lnot \bot^{BC}_A}
\newcommand{\notabortCFB}{\lnot \bot^{BC}_B}

\newcommand{\notabortOTB}{\lnot \bot^{OT}_B}

\begin{document}
\setlength{\textheight}{8.0truein} 

\thispagestyle{empty}
\setcounter{page}{1}

\title{\bf Lower bounds for quantum oblivious transfer}

\author{%
Andr\'e Chailloux\thanks{%
Centrum Wiskunde Informitica, Amsterdam, The Netherlands. Email: {\tt 
A.G.Chailloux@cwi.nl}.}
\and
Iordanis Kerenidis\thanks{%
Laboratoire d'Informatique Algorithmique: Fondements et Applications, Universit\'e Paris Diderot. 
Email: {
\tt 
iordanis.kerenidis@liafa.univ-paris-diderot.fr}.}
\and
Jamie Sikora\thanks{%
Laboratoire d'Informatique Algorithmique: Fondements et Applications, Universit\'e Paris Diderot. 
Email: {
\tt 
jamie.sikora@liafa.univ-paris-diderot.fr}.}
}

\date{ \quad \\ March 25, 2013}                                           

\maketitle

\begin{abstract}
Oblivious transfer is a fundamental primitive in cryptography. While perfect information theoretic security is impossible, quantum oblivious transfer protocols can limit the dishonest player's cheating. Finding the optimal security parameters in such protocols is an important open question. 
In this paper we show that every 1-out-of-2 oblivious transfer protocol allows a dishonest party to cheat with probability bounded below by a constant strictly larger than $1/2$. Alice's cheating is defined as her probability of guessing Bob's index, and Bob's cheating is defined as his probability of guessing both input bits of Alice. In our proof, we relate these cheating probabilities to the cheating probabilities of a bit commitment protocol and conclude by using lower bounds on quantum bit commitment. 
Then, we present an oblivious transfer protocol with two messages and cheating probabilities at most $3/4$. Last, we extend Kitaev's semidefinite programming formulation to more general primitives, where the security is against a dishonest player trying to force the outcome of the other player, and prove optimal lower and upper bounds for them. 
\end{abstract} 

\section{Introduction}        

Quantum information enables us to do cryptography with information theoretic security. The first breakthrough result in quantum cryptography is the unconditionally secure key distribution protocol of Bennett and Brassard~\cite{BB84}. Since then, a long series of work has studied which other cryptographic primitives are possible in the quantum world. However, the subsequent results were negative. Mayers and Lo, Chau proved the impossibility of secure ideal quantum bit commitment and oblivious transfer and consequently of any type of two-party secure computation~\cite{May97, LC97, Lo97, DKSW07}. On the other hand, several imperfect variants of these primitives have been shown to be possible. Finding the optimal parameters for such fundamental primitives has been since an important open question. The reason for looking at these abstract primitives is that they are the basis for all cryptographic protocols one may wish to construct, including identification schemes, digital signatures, electronic voting, etc. Let us emphasize that in this paper we only look at information theoretic security and we do not discuss computational security or security in restricted models like the bounded-storage~\cite{DFSS08} or noisy-storage model~\cite{S10}.
 Since the only assumptions used in this paper are those imposed by the laws of quantum mechanics, finding the attainable levels of security is important to understand the physical limits of these primitives and of quantum protocols in general. These limits also provide a new way of looking at quantum mechanics from the point of view of information or as a theory that arises from the (im)possibility of certain cryptographic primitives. Needless to say, since we restrict to the study of protocols that are stand-alone and with imperfect security, they may not be well-suited for practical cryptography. 

We start with coin flipping, which was first proposed by Blum~\cite{Blu81} and has since found numerous applications in two-party secure computation. Even though the results of Mayers and of Lo and Chau exclude the possibility of perfect quantum coin flipping, i.e., where the resulting coin is perfectly unbiased, it still remained open whether one can construct a quantum protocol where no player could bias the coin with probability 1. Aharonov et al.~\cite{ATVY00} provided such a protocol where no dishonest player could bias the coin with probability higher than 0.9143. Then, Ambainis \cite{Amb01} described an improved protocol whose cheating probability was at most $3/4$. Subsequently, a number of different protocols had been proposed~\cite{SR01,NS03,KN04} that achieved the same bound of $3/4$.

On the other hand, Kitaev~\cite{Kit03}, using a formulation of quantum coin flipping as semidefinite programs proved a lower bound of $1/2$ on the product of the cheating probabilities for Alice and Bob (see \cite{ABD+04}). In other words, no quantum coin flipping protocol can achieve a cheating probability less than $1/\sqrt{2}$ for both Alice and Bob. 

The question of whether $3/4$ or $1/\sqrt{2}$ was the right answer has recently been resolved by Chailloux and Kerenidis \cite{CK09} who described a protocol with cheating probability arbitrarily close to $1/\sqrt{2}$. In their protocol they use as a subroutine a weaker variant of coin flipping which is referred to as {\em weak coin flipping}. 

Weak coin flipping protocols with cheating probabilities less than $3/4$ were first constructed in \cite{SR02,Amb02,KN04}. The best bound was in fact $1/\sqrt{2}$ until the breakthrough result by Mochon who described a protocol with cheating probability $2/3$~\cite{Moc05} and then a protocol that achieves a cheating probability of $1/2+\eps$ for any $\eps>0$ ~\cite{Moc07}. Hence the optimal biases for weak and strong coin flipping are now known. 

The question has also been solved for quantum bit commitment where it was shown that there exist quantum bit commitment protocols with cheating probabilities arbitrarily close to $0.739$ and, moreover, that such protocols are optimal~\cite{CK11}.

In this paper, we focus on oblivious transfer \cite{R81,EGL82,C87}, which is a universal primitive for any two-party secure computation \cite{Kil88}. We define a 1-out-of-2 random oblivious transfer protocol with bias $\varepsilon$, denoted here as random-$OT$, to be a protocol where:
\begin{itemize}
\item Alice outputs two uniformly random bits $(x_0, x_1)$
\item Bob outputs $x_b$ and $b$ for a uniformly random bit $b$
\item $A_{OT} := \sup\{\Pr[\text{Alice guesses } b \mbox{ and Bob does not abort } ]\} = \frac{1}{2} + \varepsilon_A$
\item $B_{OT} := \sup\{\Pr[\text{Bob guesses } (x_0, x_1) \mbox{ and Alice does not abort }]\} = \frac{1}{2} + \varepsilon_B$
\item The bias of the protocol is defined as $\eps := \max\{\eps_A,\eps_B\}$
\end{itemize}
where the suprema are taken over all strategies for Alice and Bob respectively.  Note that in our definition, the bias is not defined just as an upper bound on the cheating probabilities but corresponds to the optimal cheating probability.

We note here that an honest Bob can learn both bits with probability $1/2$, since he can learn one bit perfectly and can make a random guess for the other bit. 

There is also another variant, denoted as $OT$, where Alice and Bob have specific values of $(x_0,x_1)$ and $b$ as inputs. We show that the two notions are equivalent with respect to $\varepsilon$.

The first impossibility result for quantum $OT$ with information theoretic security was shown by Lo \cite{Lo97}. The main idea is that if Alice has no information about Bob's index $b$ then Bob can learn both bits in the following way: first, Bob honestly runs the protocol with $b=0$ to learn $x_0$ with probability 1; then he locally applies a unitary to his part of the joint final state in order to transform the joint state to the joint final state in the case of $b=1$ and hence learn $x_1$. Since, Bob can learn each bit with probability 1, his measurement does not change the state and hence he can perform both of them sequentially.   

However, not much was known about the best possible bias that one can get for $OT$. In high level, $OT$ is the ``strongest'' primitive, since it implies bit commitment, coin flipping, and in fact any two-party functionality. However, when one looks at the optimal constant values for the bias, then one needs to be more careful. For example, the standard way of constructing a bit commitment protocol from $OT$ is the following: Alice and Bob perform $OT$ with inputs $x_0,x_1$, where $x_0 \oplus x_1$ is the committed bit. Since Bob can learn only one of the two inputs, he has no information about the committed bit. On the other hand, in the reveal phase, Alice reveals both bits, and since she has no information about which one Bob has learnt, if she wants to change her mind without getting caught, she can only do it with probability $1/2$ (hence her cheating probability is $3/4$). Classically, one can then repeat this protocol many times in order to take this probability close to $1/2$. As we can see, a perfect $OT$ protocol does not automatically give a perfect bit commitment protocol, as there is a loss in the parameters. Hence, Kitaev's lower bound does not a priori hold for $OT$, since we do not know how to easily convert an $OT$ protocol to a coin flipping protocol or bit commitment protocol without any loss.

Let us also note that in the quantum setting, one can use a large number of bit commitment protocols in order to construct an $OT$ protocol, something which is not known to be possible classically (\cite{Yao95},\cite{BF10}).

In related work, Salvail, Schaffner and Sotakova \cite{SSS09} have quantitatively studied a different notion of security for $OT$ protocols (and generally any two-party protocols) that they call information leakage. Information leakage  is defined as the maximum amount of extra information about the other party's output given the quantum state held by one party. They prove, among other results, that any 1-out-of-2 $OT$ protocol has a constant leakage. Their model is somewhat different, for example they do not allow the players to abort during the protocol, and their security notion is described in terms of mutual information and entropy and does not immediately translate to our security notion of guessing probabilities. However, their results provide more evidence that almost-perfect $OT$ protocols are impossible for different variants of security.

In another work, Jain, Radhakrishnan and Sen \cite{JRS02} showed that in a 1-out-of-$n$ $OT$ protocol, if Alice gets $t$ bits of information about Bob's index $b$, then Bob gets at least $\Omega(n/2^{O(t)})$ bits of information about Alice's string $x$.

In this paper, we quantitatively study the bias of quantum oblivious transfer protocols. More precisely, we construct a bit commitment protocol that uses $OT$ as a subroutine and show a relation between the cheating probabilities of the $OT$ protocol and the ones of the bit commitment protocol. Then, using a lower bound for quantum bit commitment, we derive a non-trivial lower bound (albeit weaker) on the cheating probabilities for $OT$. More precisely we prove the following theorem.

\vspace*{12pt}
\noindent
{\bf Theorem~1:} 
In any quantum oblivious transfer protocol, we have
\begin{equation*}
\max \set{A_{OT},B_{OT}} \ge 0.5852.
\end{equation*}
\vspace*{0pt}

This result shows that there is a constant lower bound on the bias of every $OT$ protocol. In other words, the bias cannot be ``amplified'' to the level used in completeness assumptions. Thus, further restrictions (beyond those enforced by the laws of quantum mechanics) are needed to use quantum $OT$ protocols as a ``building block'' for more elaborate protocols.

In Section \ref{UpperBoundSection}, we describe a simple 1-out-of-2 random-$OT$ protocol and analyze the cheating probabilities of Alice and Bob.

\vspace*{12pt}
\noindent
{\bf Theorem~2:}
There exists a quantum oblivious transfer protocol such that
$ A_{OT} = B_{OT} = \frac{3}{4}$.
\vspace*{12pt}

One may wonder if it would be possible to extend Kitaev's semidefinite programming formulation to include the $OT$ primitive and get a stronger lower bound this way. In fact, in Section~\ref{forcing} we describe a generalisation of Kitaev's semidefinite program that captures a variant of the general $k$-out-of-$n$ $OT$ primitive. Coin flipping, is then the special case of 1-out-of-1 $OT$. However, there is a big difference. What the semidefinite programming formulation captures is the probability that one party can force the outcome of the other party. 

More precisely, we define a $k$-out-of-$n$ forcing oblivious transfer protocol, denoted here as $\knfOT$, with \emph{forcing bias} $\varepsilon$ as a protocol satisfying:
\begin{itemize}
\item Alice outputs $n$ random bits $x := (x_1, \ldots, x_n)$
\item Bob outputs a random index set $b$ of $k$ indices and bit string $x_b$ consisting of $x_i$ for $i \in b$
\item $A_{b, x_b} := \sup\{\Pr[\text{Alice can force Bob to output } (b, x_b)]\} = \dfrac{\varepsilon_A}{\chews{n}{k} \cdot 2^k}$
\item $B_x := \sup\{\Pr[\text{Bob can force Alice to output } x]\} = \dfrac{\varepsilon_B}{2^n}$
\item The forcing bias of the protocol is defined as $\eps := \max\{\eps_A,\eps_B\}$
\end{itemize}
where, again, the suprema are over all strategies of Alice and Bob respectively.
First, notice our definition of the bias $\varepsilon$ as a multiplicative factor instead of additive. We choose this since the honest probabilities of the two players can be very different and in this case our definition makes more sense.

More importantly, this ``forcing'' security definition is exactly what is needed in coin flipping since, there, Alice and Bob know each others outputs and the only cheating is forcing the other player's output in order to get a specific value for the coin. However, this is very different than the probability that one player can guess the outcome of the other player, which is the security guarantee we wish for in an $OT$ protocol. 
 Due to this subtle difference in the security definition, there are few connections between the bias of an $OT$ protocol and to that of its forcing variant. For example, a strategy for Bob which allows him to \emph{force} Alice to output $(x_0, x_1) = (0,0)$ with probability $1/2$ is nontrivial in the forcing setting. However, the same strategy may only let Bob \emph{learn} $(x_0, x_1)$ with probability $1/2$, which is the same probability had he been honest and learned $x_0$ or $x_1$ perfectly.

Nevertheless, it is still interesting to know how one can extend Kitaev's semidefinite programming formulation, what are the most general primitives that can be described in this framework, and what are their applications. For these $k$-out-of-$n$ ``forcing'' primitives we provide optimal upper and lower bounds.

\vspace*{12pt}
\noindent
{\bf Theorem~3:}
 In any $\knfOT$ protocol and consistent $b,x,x_b$ (that is, when $x_b$ is the $b$ entries of $x$) we have 
\begin{equation*}
B_{{x}} \cdot A_{{b}, {x_b}} \geq \Pr[\text{Alice honestly outputs } {x} \text{ and Bob honestly outputs } ({b}, {x_b})] = \dfrac{1}{\chews{n}{k} 2^n}. 
\end{equation*}
In particular, the forcing bias satisfies $\varepsilon \geq \sqrt{2}^k$. 
\vspace*{12pt}

Note that for the special case of coin flipping, which is $\fOT{1}{1}$, our bounds are tight (a multiplicative bias of $\sqrt{2}$ is equivalent to a cheating probability of $\frac{1}{\sqrt{2}}$).

Similar to coin flipping, one can get optimal protocols as well for $\knfOT$.

\vspace*{12pt}
\noindent
{\bf Theorem~4:} 
Let $\gamma>0$. There exists a protocol for $\knfOT$ with cheating probabilities:
\begin{equation*}
A_{b, x_b} \leq \dfrac{\sqrt{2}^k (1 + \gamma)}{\chews{n}{k} \cdot 2^k} \quad \mbox{and} \quad B_x  \leq \dfrac{\sqrt{2}^k (1 + \gamma)}{2^n}.
\end{equation*} 
\vspace*{0pt}

\section{Preliminaries}

\subsection{Definitions of Primitives}

In the literature, many different variants of oblivious transfer have been considered. In this paper, we consider two variants of quantum oblivious transfer and for completeness we show that they are equivalent with respect to the bias $\eps$.

\vspace*{12pt}
\noindent
{\bf Definition~5:}
[Random Oblivious Transfer]
A 1-out-of-2 quantum random oblivious transfer protocol with bias $\eps$, denoted here as random-$OT$, is a protocol between Alice and Bob such that:
\begin{itemize}
\item Alice outputs two bits $(x_0,x_1)$ or Abort and Bob outputs two bits $(b,y)$ or Abort
\item If Alice and Bob are honest, they never Abort, $y = x_b$, Alice has no information about $b$ and Bob has no information about $x_{\overline{b}}$. Also, $x_0,x_1,b$ are uniformly random bits.
\item $A_{OT} := \sup\{\Pr[\text{Alice guesses } b \text{ and Bob does not Abort}] \}= \frac{1}{2} + \eps_A$
\item $B_{OT} := \sup\{\Pr[\text{Bob guesses } (x_0, x_1) \text{ and Alice does not Abort}]\} = \frac{1}{2} + \eps_B$
\item The bias of the protocol is defined as $\eps := \max\{\eps_A,\eps_B\}$
\end{itemize}
where the suprema are taken over all cheating strategies for Alice and Bob. 
\vspace*{12pt}

Note that this definition is slightly different from usual definitions because we want the exact value of the cheating probabilities and not only an upper bound. This is because we consider both lower bounds and upper bounds for $OT$ protocols but we could have equivalent results using the standard definitions.

An important issue is that we quantify the security against a cheating Bob as the probability that he can guess $(x_0,x_1)$. One can imagine a security definition where Bob's guessing probability is not for $(x_0,x_1)$, but for example for $x_0 \oplus x_1$ or any other function $f(x_0,x_1)$. Since we are mostly interested in lower bounds, we believe our definition is the most appropriate one, since a lower bound on the probability of guessing $(x_0,x_1)$ automatically yields a lower bound on the probability of guessing any $f(x_0,x_1)$. Note that in classical cryptography, if Bob does not know $x_0 \oplus x_1$, then this implies he does not know one of the two bits \cite{DFSS06}.

We now define a second notion of $OT$ where the values $(x_0,x_1)$ and $b$ are Alice's and Bob's inputs respectively and show the equivalence between the two notions.

\vspace*{12pt}
\noindent
{\bf Definition~6:}
[Oblivious Transfer]
A 1-out-of-2 quantum oblivious transfer protocol with bias $\eps$, denoted here as $OT$, is a protocol between Alice and Bob such that:
\begin{itemize}
\item Alice has input $x_0,x_1 \in \zo$ and Bob has input $b \in \zo$. At the beginning of the protocol, Alice has no information about $b$ and Bob has no information about $(x_0,x_1)$
\item At the end of the protocol, Bob outputs $y$ or Abort and Alice can either Abort or not
\item If Alice and Bob are honest, they never Abort, $y = x_b$, Alice has no information about $b$ and Bob has no information about $x_{\overline{b}}$
\item $A_{OT} := \sup\{\Pr[\text{Alice guesses } b \text{ and Bob does not Abort}]\} = \frac{1}{2} + \eps_A$
\item $B_{OT} := \sup\{\Pr[\text{Bob guesses } (x_0, x_1) \text{ and Alice does not Abort}]\} = \frac{1}{2} + \eps_B$
\item The bias of the protocol is defined as $\eps := \max\{\eps_A,\eps_B\}$
\end{itemize}
where the suprema are taken over all cheating strategies for Alice and Bob. 
\vspace*{12pt}

We also define quantum bit commitment.

\vspace*{12pt}
\noindent
{\bf Definition~7:}
[Bit Commitment]
A quantum commitment scheme  is an interactive protocol between Alice and Bob with two phases, a Commit phase and a Reveal phase.
\begin{itemize}
\item In the {\em Commit} phase, Alice interacts with Bob   in order to commit to $b$
\item In the {\em Reveal} phase, Alice interacts with Bob in order to reveal $b$. Bob decides to accept or reject depending on the revealed value of $b$ and his final state. We say that Alice successfully reveals $b$ if Bob accepts the revealed value.
\end{itemize} 
\noindent We define the following security requirements for the commitment scheme.
\begin{itemize}
\item \emph{Completeness:} If Alice and Bob are both honest then Alice always successfully reveals the bit $b$ which she has committed
\item  \emph{Binding property:} For any cheating Alice and for honest Bob, we define Alice's cheating probability as
\begin{equation*}
A_{BC} = \frac{1}{2}\left(
\Pr[\mbox{Alice successfully reveals } b = 0 ] + 
\Pr[\mbox{Alice successfully reveals } b = 1 ]\right)
\end{equation*}
\item \emph{Hiding property:} For any cheating Bob and for honest Alice, we define Bob's cheating probability as
\begin{equation*}
B_{BC} = 
\Pr[\mbox{Bob guesses } b \mbox{ after the Commit phase}]. 
\end{equation*}
\end{itemize} 
\vspace*{0pt}

\paragraph{Remark:} 

The definition of quantum bit commitment we use is the standard one when one studies stand-alone cryptographic primitives. In this setting, quantum bit commitment has a clear relation to other fundamental primitives such as coin flipping and oblivious transfer~\cite{ATVY00,Amb01,Kit03,Moc07}. Moreover, the study of such primitives sheds light on the physical limits of quantum mechanics and the power of entanglement. Recently there have been some stronger definitions of quantum bit commitment protocols that suit better practical uses (see for example~\cite{DFR+07}). 

We have the following lower bound for quantum bit commitment~\cite{CK11}.

\vspace*{12pt}
\noindent
{\bf Proposition~8:}
[\cite{CK11}] 
For any quantum bit commitment scheme with cheating probabilities $A_{BC}$ and $B_{BC}$, there is a parameter $t \in [0,1]$ such that 
\begin{equation*}
A_{BC} \ge \left( 1 - \left( 1 - \frac{1}{\sqrt{2}} \right)t \right)^2 \quad \text{ and } \quad B_{BC} \ge \frac{1}{2} + \frac{t}{2}.
\end{equation*} 
\vspace*{0pt}

\subsection{Equivalence Between the Different Notions of Oblivious Transfer}
We show the equivalence between $OT$ and random-$OT$ with respect to the bias $\eps$.

\vspace*{12pt}
\noindent
{\bf Proposition~9:}
Let $P$ be an $OT$ protocol with bias $\eps$. We can construct a random-$OT$ protocol $Q$ with bias $\eps$ using $P$. 

\vspace*{12pt}
\noindent
{\bf Proof:}
The construction of the $OT$ protocol $Q$ is pretty straightforward:
\begin{enumerate}
\item Alice picks $x_0,x_1 \in_R \zo$ uniformly at random and Bob picks $b \in_R \zo$ uniformly at random.
\item Alice and Bob perform the $OT$ protocol $P$ where Alice inputs $x_0,x_1$ and Bob inputs $b$. Let $y$ be Bob's output. Note that at this point, Alice has no information about $b$ and Bob has no information about $(x_0,x_1)$.
\item Alice and Bob abort in $Q$ if and only if they abort in $P$. Otherwise, the outputs of protocol $Q$ are $(x_0,x_1)$ for Alice and $(b,y)$ for Bob.
\end{enumerate}
The outcomes of $Q$ are uniformly random bits since Alice and Bob choose their inputs uniformly at random. All the other requirements that make $Q$ an $OT$ protocol with bias $\eps$ are satisfied because $P$ is an $OT$ protocol with bias $\eps$. \, $\square$

\vspace*{12pt}

We now prove how to go from a random-$OT$ to an $OT$ protocol.

\vspace*{12pt}
\noindent
{\bf Proposition~10:}
Let $P$ be a random-$OT$ protocol with bias $\eps_P$. We can construct an $OT$ protocol $Q$ with bias $\eps_Q = \eps_P$ using $P$.

\vspace*{12pt}
\noindent
{\bf Proof:}
Let $P$ be a random-$OT$ protocol with bias $\eps_P$. Consider the following protocol $Q$:
\begin{enumerate}
\item Alice has inputs $X_0,X_1$ and Bob has an input $B$.
\item Alice and Bob run protocol $P$ and output $(x_0,x_1)$ for Alice and $(b,y)$ for Bob. 
\item Bob sends $r = b \oplus B$ to Alice. Let $x'_c=x_{c \oplus r}$, for $c\in \{0,1\}$. 
\item Alice sends to Bob $(s_0,s_1)$ where $s_c = x'_c \oplus X_c$ for $c \in \zo$. Let $y' = y \oplus s_{B}$.
\item Alice and Bob abort in $Q$ if and only if they abort in $P$. Otherwise, the output of the protocol is $y'$ for Bob. 
\end{enumerate}
We now show that our protocol is an $OT$ protocol with bias $\eps_P$. 
First, note that the values $x'_c$ are known by Alice and the value $y'$ is known by Bob. Also, notice that $x'_{B} = x_{B \oplus r} = x_b$.
\begin{itemize}
\item Alice and Bob are honest: 
By definition we have $y = x_b$. Then, we have 
\begin{equation*} 
y' = y \oplus s_{B} = x_{b} \oplus s_{B} = x'_{B} \oplus s_{B} = X_B.
\end{equation*} 
Moreover,
Alice knows $r$ but has no information about $b$ and hence she has no information about $B = b \oplus r$. Bob knows $(s_0,s_1)$ and $r$ but has no information about $x_{\bar{b}}$, hence he has no information about $X_{\bar{B}} = x'_{\bar{B}} \oplus s_{\bar{B}} = x'_{\bar{b} \oplus r} \oplus s_{\bar{b} \oplus r} = x_{\bar{b}} \oplus s_{\bar{b} \oplus r}$. 

\item Cheating Alice: 
Alice knows $r$ and $B = b \oplus r$. Hence
\begin{eqnarray*}
A_{OT}(Q) & = & \sup \{\Pr[\text{Alice guesses } B \text{ and Bob does not Abort}]\} \nonumber \\
& = & \sup \{\Pr[\text{Alice guesses } b \text{ and Bob does not Abort}]\} = A_{OT}(P). 
\end{eqnarray*}
\item Cheating Bob: Bob knows $r$ and $(s_0,s_1)$. We have $X_c= x'_c \oplus s_c = x_{c \oplus r} \oplus s_c$ so it is equivalent for Bob to guess $(X_0,X_1)$ and $(x_0,x_1)$. Hence
\begin{eqnarray*} 
B_{OT}(Q) & = & \sup \{ \Pr[\text{Bob guesses } (X_0,X_1) \text{ and Alice does not Abort}]\} \nonumber \\ 
& = & \sup \{\Pr[\text{Bob guesses } (x_0,x_1) \text{ and Alice does not Abort}] \} = B_{OT}(P). 
\end{eqnarray*}
\end{itemize}
We can now conclude for the biases
\begin{equation*}
\eps_Q = \max\{A_{OT}(Q),B_{OT}(Q)\} - \frac{1}{2} = \max\{A_{OT}(P),B_{OT}(P)\} - \frac{1}{2} = \eps_P. \quad \square
\end{equation*} 

\section{A Lower Bound on Any Oblivious Transfer Protocol}

In this section we prove that the bias of any random-$OT$ protocol, and hence any $OT$ protocol, is bounded below by a constant. We start from a random-$OT$ protocol and first show how to construct a bit commitment protocol. Then, we prove a relation between the cheating probabilities of the bit commitment protocol and those in the random-$OT$ protocol. Last, we use the lower bound for quantum bit commitment (Proposition~8) to derive a lower bound on any $OT$ protocol.

We create a bit commitment protocol from a random-$OT$ protocol as follows.

\vspace*{12pt}
{ \center
\begin{tabular}{|l|}
\hline \\
\quad \quad \underline{Bit Commitment Protocol via random-$OT$ $\phantom{\frac{1}{2}}$} \\
\quad \\
\quad $1.$ \emph{Commit phase}: We invert the roles of Alice and Bob. Bob is the one who commits. \qquad \\
\quad $\phantom{1.}$ He wants  to commit to a bit $a$. Alice and Bob perform the $OT$ protocol such that \qquad \\
\quad $\phantom{1.}$ Alice has $(x_0, x_1)$ and Bob has $(b, x_b)$. Bob sends $c := a \oplus b$ to Alice. \qquad \\
\quad \\
\quad $2.$ \emph{Reveal phase}: Bob reveals $b$ and $y = x_b$ to Alice. 
If $x_b$ from Bob is consistent with \qquad \\
\quad $\phantom{2.}$ Alice's bits then Alice accepts $c \oplus b = a$. Otherwise Alice aborts. \qquad \\ 
\quad \\
\hline
\end{tabular}
\quad \\
}
\vspace*{12pt}

We now analyze how much Alice and Bob can cheat in the bit commitment protocol and compare these quantities to the bias of the random-$OT$ protocol. Let $A_{OT},B_{OT}$ be the cheating probabilities for the quantum oblivious transfer protocol and $A_{BC},B_{BC}$ be the cheating probabilities for the resulting quantum bit commitment protocol. Our goal is to show the following:

\vspace*{12pt}
\noindent
{\bf Proposition~11:}
For the protocol above, we have
\begin{equation*}
A_{OT} = A_{BC} \quad \text{ and } \quad B_{OT} \ge f(B_{BC}) \;  \textrm{ where } \; f(x) = x(2x-1)^2.
\end{equation*} 

\vspace*{12pt}
\noindent
{\bf Proof:}
Let $\notabortCFA$ (resp. $\notabortCFB$) denote the event ``Alice (resp. Bob) does not abort during the entire bit commitment protocol''. Let $\notabortOTA$ (resp. $\notabortOTB$) denote the event ``Alice (resp. Bob) does not abort during the random-$OT$ subroutine''.

\vspace*{12pt}
\noindent 
{\underline{Cheating Alice}}

By definition, $A_{OT}$ is the optimal probability of Alice guessing $b$ in the random-$OT$ protocol without Bob aborting and $A_{BC}$ is the optimal probability of Alice guessing $a$ in the bit commitment protocol without Bob aborting. Since Alice knows $c := a \oplus b$, the probability of Alice guessing $a$ in the bit commitment protocol is the same as the probability of her guessing $b$ in the random-$OT$ protocol. Thus $A_{OT} = A_{BC}$.

\vspace*{12pt}
\noindent 
{\underline{Cheating Bob}}

By definition, $B_{OT}$ is the optimal probability of Bob learning both bits in the random-$OT$ protocol without Alice aborting. Thus,
\begin{eqnarray*} B_{OT} & = & \sup\{\Pr[ \mbox{ (Bob guesses } (x_0,x_1) )  \land \notabortOTA] \} \nonumber \\ 
& = &
\sup\{\Pr[ \notabortOTA ] \cdot \Pr [ \mbox{ (Bob guesses } (x_0,x_1)) | \notabortOTA]\}. \end{eqnarray*}
where the suprema are taken over all strategies for Bob.

If Bob wants to reveal $0$ in the bit commitment protocol  (a similar argument works if he wants to reveal $1$), then first, Alice must not abort in the random-$OT$ protocol and second, Bob must send $b = c$ as well as the correct $x_c$ such that Alice does not abort in the last round of the bit commitment protocol. This is equivalent to saying that Bob succeeds if he guesses $x_c$ and Alice does not abort in the random-$OT$ protocol. Since Bob randomly chooses which bit he wants to reveal, we can write the probability of Bob cheating as 
\begin{eqnarray*} 
\! B_{BC} & = & \max \left\{ \frac{1}{2} \Pr[\textrm{(Bob guesses } x_0) \land \notabortOTA] + \frac{1}{2} \Pr[\textrm{(Bob guesses } x_1) \land \notabortOTA] \right\} \nonumber \\
& = & \max \left\{ \frac{\Pr[\notabortOTA]}{2} \left( \Pr[\textrm{(Bob guesses } x_0) | \notabortOTA] + \Pr[\textrm{(Bob guesses } x_1) | \notabortOTA] \right) \right\} \! . 
\end{eqnarray*}

Notice that we use ``max'' instead of ``sup'' above. This is because an optimal strategy exists for every bit commitment protocol. To see this, we can construct a coin flipping protocol from any bit commitment protocol and an optimal strategy always exists for any coin flipping protocol. This is a consequence of strong duality in the semidefinite programming formalism of coin flipping protocols \cite{Kit03}, see \cite{ABD+04} for details.

Let us now fix Bob's optimal cheating strategy in the bit commitment protocol. For this strategy, let $p = \Pr[\textrm{(Bob guesses } x_0) | \notabortOTA]$, $q = \Pr[\textrm{(Bob guesses } x_1) | \notabortOTA]$ and let $a = \frac{p + q}{2}$. Note that, without loss of generality, we can assume that Bob's measurements are projective measurements. This can be done by increasing the dimension of Bob's space. Also, Alice has a projective measurement on her space to determine the bits $(x_0,x_1)$. 

We use the following lemma to relate $B_{BC}$ and $B_{OT}$.

\vspace*{12pt}
\noindent
{\bf Lemma~12:}
[Learning-In-Sequence Lemma] 
Let $p,q \in [1/2,1]$. Let Alice and Bob share a joint pure state. Suppose Alice performs a projective measurement $M = \{M_{x_0,x_1}\}_{x_0,x_1 \in \zo}$ on her space to determine the values of $(x_0,x_1)$. Suppose there is a projective measurement $P= \{P_0,P_1\}$ on Bob's space that allows him to guess bit $x_0$ with probability $p$ and a projective measurement $Q = \{Q_0,Q_1\}$ on his space that allows him to guess bit $x_1$ with probability $q$. Then, there exists a measurement on Bob's space that allows him to guess $(x_0,x_1)$ with probability at least $a(2a - 1)^2$ where $a = \frac{p + q}{2}$. 
\vspace*{0pt}

\vspace*{12pt}

We postpone the proof of this lemma to Subsection \ref{proof}. 

We now construct a cheating strategy for Bob for the $OT$ protocol: Run the optimal bit commitment strategy and look at Bob's state after Step $1$ conditioned on $\notabortOTA$. Note that this event happens with nonzero probability in the optimal bit commitment strategy since otherwise the success probability is $0$. The optimal bit commitment strategy gives measurements that allow Bob to guess $x_0$ with probability $p$ and $x_1$ with probability $q$. Bob uses these measurements and the procedure of Lemma~12 to guess $(x_0,x_1)$. Let $m$ be the probability he guesses $(x_0,x_1)$. From Lemma~12, we have that $m \ge a(2a-1)^2$. By definition of $B_{OT}$ and $B_{BC}$, we have:
\begin{equation*}
m = \Pr [ \mbox{ (Bob guesses } (x_0,x_1)) | \notabortOTA] \le \dfrac{B_{OT}}{\Pr[\notabortOTA]} \quad \mbox{ and } \quad a = \frac{B_{BC}}{\Pr[\notabortOTA]}. 
\end{equation*}
This gives us
\begin{equation*}
\dfrac{B_{OT}}{\Pr[\notabortOTA]} \geq \dfrac{B_{BC}}{\Pr[\notabortOTA]} \left(2 \dfrac{B_{BC}}{\Pr[\notabortOTA]} - 1\right)^2 \implies B_{OT} \geq B_{BC} \left( 2B_{BC} - 1 \right)^2, 
\end{equation*}
where the implication holds since $B_{BC} \geq 1/2$. \, $\square$ 

\vspace*{12pt}

Using this proposition and the lower bound for quantum bit commitment, we can show a lower bound.

\vspace*{12pt}
\noindent
{\bf Theorem~1:} 
In any quantum oblivious transfer protocol, at least one of the players can cheat with probability $0.5852$. 

\vspace*{12pt}
\noindent
{\bf Proof:}
We use $
A_{BC} = A_{OT}$ and $B_{OT} \ge f(B_{BC})$ (where $f(x) = x(2x-1)^2$) from
Proposition~11. From Proposition~8, we have that for any quantum bit commitment scheme, there exists a parameter $t \in [0,1]$ such that
\begin{equation*}
B_{BC} \ge \left( 1 - \left( 1-\frac{1}{\sqrt{2}} \right) t \right)^2 \ge \dfrac{1}{2} \quad \text{ and } \quad A_{BC} \ge \frac{1}{2} +\frac{t}{2} \ge \dfrac{1}{2},
\end{equation*}
noting that we have reversed the roles of Alice and Bob in the bit commitment protocol.
We immediately have that there exists a parameter $t\in [0,1]$ such that
\begin{equation*}
B_{OT} \ge f \left( 1 - \left( 1-\frac{1}{\sqrt{2}} \right) t \right)^2 \quad \text{ and } \quad A_{OT} \ge \frac{1}{2} +\frac{t}{2}, 
\end{equation*}
since $f$ is nondecreasing on the interval $[1/2, 1]$.
We get a lower bound on $\max\set{A_{OT}, B_{OT}}$ by equating the lower bounds above and solving for $t \approx 0.1705$ ($t$ is the solution of a degree 6 polynomial, thus we do not know any closed form for $t$). At this value of $t$, we have $A_{OT}, B_{OT} \approx 0.5852$, yielding the desired bound. \, $\square$ 

\subsection{Proof of the Learning-In-Sequence Lemma} \label{proof}

We first provide three claims that enable us to prove  a geometric statement about performing two projective measurements in sequence on the same quantum state (Lemma~16) and then, we use this geometric lemma to prove the Learning-in-Sequence Lemma.

\vspace*{12pt}
\noindent
{\bf Claim~13:}
Let $\ket{X}$ and $\ket{Y}$ be pure states and $Q$ a projection such that $Q\ket{Y} = \ket{Y}$. Then we have
\begin{equation*}
\| Q\ket{X} \|_2^2 \ge |\braket{X}{Y}|^2. 
\end{equation*}

\vspace*{12pt}
\noindent
{\bf Proof:}
Using Cauchy-Schwarz we have
\begin{equation*}
| \braket{X}{Y} |^2 = | \bra{X}Q  \ket{ Y} |^2 \leq \norm{Q \ket{X}}_2^2 \norm{\ket{Y}}_2^2 = \norm{Q \ket{X}}_2^2. \quad \square 
\end{equation*}

\vspace*{12pt}
\noindent
{\bf Claim~14:}
Suppose $\theta, \theta' \in [0,\pi/4]$. If $| \braket{\psi}{\phi} | \ge \cos(\theta)$ and $| \braket{\phi}{\xi} | \ge \cos(\theta')$ then 
\begin{equation*}
| \braket{\psi}{\xi} | \ge \cos(\theta + \theta'). 
\end{equation*}
 
\vspace*{12pt}
\noindent
{\bf Proof:}
Define the angle between two pure states $\ket{\psi}$ and $\ket{\phi}$ as $A(\psi, \phi) := \arccos | \braket{\psi}{\phi} |$. This is a metric (see~\cite{NC00} page 413). Thus we have 
\begin{equation*} 
\arccos | \braket{\psi}{\xi} | = A(\psi, \xi) \leq A(\psi, \phi) + A(\phi, \xi) = \arccos | \braket{\psi}{\phi} | + \arccos | \braket{\phi}{\xi} | \leq \theta + \theta'. 
\end{equation*}
Taking the cosine of both sides yields the result. \, $\square$ 

\vspace*{12pt}
\noindent
{\bf Claim~15:}
Let $\theta, \theta' \in [0,\pi/4]$. Then
\begin{equation*} 
\cos(\theta + \theta') \ge \cos^2(\theta) + \cos^2(\theta') - 1. 
\end{equation*} 

\vspace*{12pt}
\noindent
{\bf Proof:}
Without loss of generality suppose that $\theta \ge \theta'$. Consider the function
\begin{equation*} 
f(\theta) = \cos(\theta + \theta') - \cos^2(\theta) + \sin^2(\theta') 
\end{equation*} 
for fixed $\theta'$. Taking its derivative we get
\begin{equation*} 
f'(\theta) = -\sin(\theta + \theta') + \sin(2\theta) 
\end{equation*}
which is nonnegative for $\theta \in [\theta', \pi/4]$. Since $f(\theta') = 0$, we conclude that $f(\theta) \ge 0$ for $\theta \in[\theta', \pi/4]$ which gives the desired result. \, $\square$ 

\vspace*{12pt}
\noindent
{\bf Lemma~16:} 
Let $\ket{\psi}$ be a pure state and let $\{C, I-C\}$ and $\{D, I-D\}$ be two projective measurements such that
\begin{equation*} 
\cos^2(\theta) := \norm{C \ket{\psi}}^2_2 \geq \half 
\quad \text{ and } \quad 
\cos^2(\theta') := \norm{D \ket{\psi}}^2_2 \geq \half. 
\end{equation*}
Then we have
\begin{equation*}
\norm{DC \ket{\psi} }_2^2 \ge \cos^2(\theta) \cos^2(\theta + \theta'). 
\end{equation*}

\vspace*{12pt}
\noindent
{\bf Proof:}
Define the following states
\begin{equation*} 
\ket{X} := \dfrac{C \ket{\psi}}{\norm{C \ket{\psi}}_2}, \quad
\ket{X'} := \dfrac{(I-C) \ket{\psi}}{\norm{(I-C) \ket{\psi}}_2}, \quad
\ket{Y} := \dfrac{D \ket{\psi}}{\norm{D \ket{\psi}}_2}, \quad 
\ket{Y'} := \dfrac{(I-D) \ket{\psi}}{\norm{(I-D) \ket{\psi}}_2}. 
\end{equation*}
Then we can write
$ \ket{\psi} = \cos(\theta)\ket{X} + e^{i\alpha}\sin(\theta)\ket{X'} 
\; \text{ and } \;
\ket{\psi} = \cos(\theta')\ket{Y} + e^{i \beta} \sin(\theta')\ket{Y'}$
with $\alpha,\beta \in \R$.
Then we have
\begin{eqnarray*}
\norm{DC \ket{\psi}}_2^2 
& = & \cos^2(\theta) \norm{D \ket{X}}_2^2 \nonumber \\
& \ge & \cos^2(\theta)|\braket{Y}{X}|^2 \;\; \textrm{(Claim~13)} \nonumber \\
& \ge & \cos^2(\theta)\cos^2(\theta + \theta') \;\; \textrm{(Claim~14)}. \quad \square 
\end{eqnarray*}

We now prove Lemma~12.

\vspace*{12pt}
\noindent
{\bf Proof:}
Let $\ket{\Omega}_{\spa{A}\spa{B}}$ be the joint pure state shared by Alice and Bob, where $\spa{A}$ is the space controlled by Alice and $\spa{B}$ the space controlled by Bob.

Let $M = \{M_{x_0,x_1}\}_{x_0,x_1 \in \zo}$ be Alice's projective measurement on $\spa{A}$ to determine her outputs $x_0,x_1$.  Let $P = \{P_0,P_1\}$ be Bob's projective measurement that allows him to guess $x_0$ with probability $p = \cos^2(\theta)$ and $Q = \{Q_0,Q_1\}$ be Bob's projective measurement that allows him to guess $x_1$ with probability $q = \cos^2(\theta')$.  These measurements are on $\spa{B}$ only. Now we can write $a = \frac{p+q}{2} = \frac{\cos^2(\theta) + \cos^2(\theta')}{2}$. We consider the following projections on $\spa{A}\spa{B}$:
\begin{equation*} 
C = \sum_{x_0,x_1} M_{x_0,x_1} \otimes P_{x_0} \quad \text{ and } \quad D = \sum_{x_0,x_1} M_{x_0,x_1} \otimes Q_{x_1}. 
\end{equation*}

$C$ (resp. $D$) is the projection on the subspace where Bob guesses correctly the first bit (resp. the second bit) after applying $P$ (resp. $Q$).

A strategy for Bob to learn both bits is simple: apply the two measurements $P$ and $Q$ one after the other, where the first one is chosen uniformly at random.

The projection on the subspace where Bob guesses  $(x_0,x_1)$ when applying $P$ then $Q$ is
\begin{equation*} 
E = \sum_{x_0,x_1} M_{x_0,x_1} \otimes Q_{x_1}P_{x_0} = DC. 
\end{equation*}
Similarly, the projection on the subspace where Bob guesses $(x_0,x_1)$ when applying $Q$ then $P$ is 
\begin{equation*} 
F = \sum_{x_0,x_1} M_{x_0,x_1} \otimes P_{x_0}Q_{x_1} = CD. 
\end{equation*}
With this strategy Bob can guess both bits with probability 
\begin{eqnarray*} 
\frac{1}{2}\left(||E \ket{\Omega}||^2_2 + ||F \ket{\Omega}||_2^2\right) 
& = & 
\frac{1}{2}\left(||D C \ket{\Omega}||^2_2 + ||C D \ket{\Omega}||_2^2\right) \nonumber \\
& \ge & \frac{1}{2}\left(\cos^2(\theta) + \cos^2(\theta')\right)\cos^2(\theta + \theta') \;\; \textrm{(Lemma~16)} \nonumber \\
& \ge & \frac{1}{2}\left(\cos^2(\theta) + \cos^2(\theta')\right) \left( \cos^2(\theta) + \cos^2(\theta') - 1 \right)^2 \;\;  \textrm{(Claim~15)} \nonumber \\
& = & a(2a - 1)^2.
\end{eqnarray*}
Note that we can use Lemma~16 since Bob's optimal measurement to guess $x_0$ and $x_1$ succeeds for each bit with probability at least $1/2$. \, $\square$ 

\section{A Two-Message Protocol With Bias $1/4$}\label{UpperBoundSection}

We present in this section a random-$OT$ protocol with bias $1/4$. This also implies, as we have shown, an $OT$ protocol with inputs with the same bias. 

\vspace*{12pt}
{ \center
\quad \\
\begin{tabular}{|l|}
\hline \\
\quad \quad \underline{Random Oblivious Transfer Protocol $\phantom{\frac{1}{2}}$}
\quad \\ 
\quad \\ 
\quad $1.$ Bob chooses $b \in_R \set{0,1}$ and creates the state $\ket{\phi_b} := \frac{1}{\sqrt{2}} \ket{bb} + \frac{1}{\sqrt{2}} \ket{22}$. \\ 
\quad $\phantom{1.}$ Bob sends one of the qutrits to Alice. \\
\quad \\
\quad $2.$ Alice chooses $x_0, x_1 \in_R \set{0,1}$ and applies the unitary $\ket{a} \to (-1)^{x_a} \ket{a}$, \quad \\
\quad $\phantom{2.}$ where $x_2 := 0$, on Bob's qutrit. \quad \\
\quad \\
\quad $3.$ Alice returns the qutrit to Bob who now has the state $\ket{\psi_b} := \frac{(-1)^{x_b}}{\sqrt{2}} \ket{bb} + \frac{1}{\sqrt{2}} \ket{22}$. \qquad \\
\quad \\
\quad $4.$ Bob performs  the measurement $\{ \Pi_0 = \kb{\phi_b}, \; \Pi_1 := \kb{\phi_b'}, \; I - \Pi_0 - \Pi_1 \}$ \qquad \\
\quad $\phantom{4.}$ on the state $\ket{\psi_b}$, where $\ket{\phi'_b} := \frac{1}{\sqrt{2}} \ket{bb} - \frac{1}{\sqrt{2}} \ket{22}$. \qquad \\
\quad \\ 
\quad $5.$ If the outcome is $\Pi_0$ then $x_b=0$, if it is $\Pi_1$ then $x_b=1$, otherwise he aborts. \\
\quad \\
\hline
\end{tabular}
\quad \\
}
\vspace*{12pt}

It is clear that Bob can learn $x_0$ or $x_1$ perfectly. Moreover, note that if he sends half of the state $\frac{1}{\sqrt 2} \ket{00} + \frac{1}{\sqrt 2} \ket{11}$ then he can also learn $x_0 \oplus x_1$ perfectly (although in this case he does not learn either of $x_0$ or $x_1$). We now show that it is impossible for him to perfectly learn both $x_0$ and $x_1$ and also that his bit is not completely revealed to a cheating Alice.

\vspace*{12pt}
\noindent
{\bf Theorem~2:} 
In the protocol described above, we have 
$A_{OT} = B_{OT} = \frac{3}{4}$. 

\vspace*{12pt}
\noindent
{\bf Proof:}
We analyze the cheating probabilities of each party. 

\vspace*{12pt}
\noindent 
{\underline{Cheating Alice}}

Define Bob's space as $\calB$ and let $\sigma_b := \Tr_{\calB}(\kb{\phi_b})$ denote the two reduced states Alice may receive in the first message. Then, the optimal strategy for Alice to learn $b$ is to perform the optimal measurement to distinguish between $\sigma_0$ and $\sigma_1$. In this case, she succeeds with probability 
\begin{equation*} 
\frac{1}{2} + \frac{1}{4} \norm{\sigma_0 - \sigma_1}_{tr} = \frac{3}{4}, 
\end{equation*}
(see for example~\cite{KN04}). Alice's optimal measurement is, in fact, a measurement in the computational basis. If she gets outcome $\ket{0}$ or $\ket{1}$ then she knows $b$ with certainty. If she gets outcome $\ket{2}$ then she guesses. Notice also, that even after this measurement she can return the measured qutrit to Bob and the outcome of Bob's measurement will always be either $\Pi_0$ or $\Pi_1$. Hence, Bob will never abort. 

\vspace*{12pt}
\noindent 
{\underline{Cheating Bob}}

 Bob wants to learn both bits $(x_0, x_1)$. We now describe a general strategy for Bob:
\begin{itemize}
\item Bob creates $\ket{\psi} = \sum_{i} \alpha_{i} \ket{i}_{\calA} \ket{e_i}_{\calB}$ and sends the $\calA$ part to Alice. The $\ket{e_i}$'s are not necessarily orthogonal but $\sum_i |\alpha_i|^2 = 1$.
\item Alice applies $U_{x_0,x_1}$ on her part and sends it back to Bob. We can write Bob's state as $\ket{\psi_{x_0, x_1}} = \sum_{i} \alpha_{i} (-1)^{x_i} \ket{i} \ket{e_i}$ recalling that $x_2 := 0$.
\end{itemize}
At the end of the protocol, Bob applies a two-outcome measurement on $\ket{\psi_{x_0, x_1}}$ to get his guess for $(x_0,x_1)$. 

From this strategy, we create another strategy with the same cheating probability where Bob sends a pure state. We define this strategy as follows:
\begin{itemize}
\item Bob creates $\ket{\psi'} = \sum_{i} \alpha_{i} \ket{i}_{\calA}$ and sends the whole state to Alice.
\item Alice applies $U_{x_0,x_1}$ on her part and sends it back to Bob. We can write Bob's state as $\ket{\psi'_{x_0, x_1}} = \sum_{i} \alpha_{i} (-1)^{x_i} \ket{i}$ recalling that $x_2 := 0$.
\item Bob applies the unitary $U: \ket{i} \ket{0} \to \ket{i} \ket{e_i}$ to $\ket{\psi'_{x_0, x_1}}\ket{0}$ and obtains $\ket{\psi_{x_0, x_1}}$.
\end{itemize}
To determine $(x_0,x_1)$, Bob applies the same measurement as in the original strategy.

Clearly both strategies have the same success probability. When Bob uses the second strategy, Alice and Bob are unentangled after the first message and Alice sends back a qutrit to Bob. Using an information bound (reproduced below for completeness), we have
\begin{equation*} 
\Pr [\textrm{Bob correctly guesses } (x_0,x_1)] \le 3/4. 
\end{equation*}

\vspace*{12pt}
\noindent
{\bf Claim~17:}
[\cite{DW09} following \cite{Nay99}] 
Suppose we have a classical random variable $X$, uniformly distributed over $[n] = \{1,\dots,n\}$. Let $x \rightarrow \ket{\phi_x}$ be some encoding of $[n]$, where $\ket{\phi_x}$ is a pure state in a $d$-dimensional space. Let $P_1,\dots,P_n$ be the measurement operators applied for decoding; these sum to the $d$-dimensional identity operator. Then the probability of correctly decoding in case $X=x$ is
\begin{equation*} 
p_x = || P_x \ket{\phi_x} ||^2 \le \Tr(P_x).
\end{equation*} 
The expected success probability is 
\begin{equation*}
\frac{1}{n} \sum_{x = 1}^n p_x \le \frac{1}{n} \sum_{x = 1}^n \Tr(P_x) = \frac{1}{n} \Tr\left(\sum_{x = 1}^n P_x \right) = \frac{1}{n} \Tr(I) = \frac{d}{n}.
\end{equation*}
\vspace*{0pt}

Note that there is a strategy for Bob to achieve $3/4$. Bob wants to learn both bits $(x_0, x_1)$. Suppose he creates the state 
\begin{equation*} 
\ket{\psi} := \frac{1}{\sqrt{3}} \ket{0} + \frac{1}{\sqrt{3}} \ket{1} + \frac{1}{\sqrt{3}} \ket{2} 
\end{equation*}
and sends it to Alice. The state he receives is 
\begin{equation*} 
\ket{\psi_{x_0,x_1}} := \frac{1}{\sqrt{3}}(-1)^{x_0} \ket{0} + \frac{1}{\sqrt{3}} (-1)^{x_1}\ket{1} + \frac{1}{\sqrt{3}} \ket{2}. 
\end{equation*}
Then, Bob performs a projective measurement in the 4-dimensional basis 
\[ \{ \ket{\Psi_{x_0,x_1}}: x_0,x_1 \in \{0,1\}\} \] 
where
\begin{equation*} 
\ket{\Psi_{x_0,x_1}} := \frac{1}{2}(-1)^{x_0} \ket{0} + \frac{1}{2} (-1)^{x_1}\ket{1} + \frac{1}{2} \ket{2} +\frac{1}{2} (-1)^{x_0 \oplus x_1}\ket{3}. 
\end{equation*}
The probability that Bob guesses the two bits $x_0,x_1$ correctly is
\begin{equation*} 
\sum_{x_0, x_1} \frac{1}{4}
\Pr [\textrm{Bob guesses } (x_0,x_1)] = \sum_{x_0, x_1} \frac{1}{4} |\langle\Psi_{x_0,x_1}|\psi_{x_0,x_1}\rangle|^2 = \frac{3}{4}.
\end{equation*}
Note that in our protocol Alice never aborts. \, $\square$ 

\section{Oblivious Transfer as a Forcing Primitive}\label{forcing}

Here, we discuss a variant of oblivious transfer, as a generalization of coin flipping, that can be analyzed using an extension of Kitaev's semidefinite programming formalism. 

\vspace*{12pt}
\noindent
{\bf Definition~18:}
[Forcing Oblivious Transfer]
A $k$-out-of-$n$ forcing oblivious transfer protocol, denoted here as $\knfOT$, with \emph{forcing bias} $\varepsilon$ is a protocol satisfying:
\begin{itemize}
\item Alice outputs $n$ random bits $x := (x_1, \ldots, x_n)$
\item Bob outputs a random index set $b$ of $k$ indices and bit string $x_b$ consisting of $x_i$ for $i \in b$
\item $A_{b, x_b} := \sup\{\Pr[\text{Alice can force Bob to output } (b, x_b)]\} = \dfrac{\varepsilon_A}{\chews{n}{k} \cdot 2^k}$
\item $B_x := \sup\{\Pr[\text{Bob can force Alice to output } x]\} = \dfrac{\varepsilon_B}{2^n}$
\item The forcing bias of the protocol is defined as $\eps=\max\{\eps_A,\eps_B\}$
\end{itemize}
where the suprema are taken over all strategies of Alice and Bob. 
\vspace*{12pt} 

The main difference in this new primitive is the definition of security. Here, we design protocols to protect against a dishonest party being able to \emph{force} a desired value as the output of the other player. In the previous section (and in the literature) oblivious transfer protocols are designed to protect against the dishonest party \emph{learning} the other party's output. Notice, for example, that in coin flipping we can design protocols to protect against a dishonest party forcing a desired outcome, but both players \emph{learn} the  coin outcome perfectly.

The primitive we have defined is indeed a generalization of coin flipping since we can cast the problem of coin flipping as a 1-out-of-1 forcing oblivious transfer protocol. Of course, in $\fOT{1}{1}$ Alice always knows Bob's index set so the forcing bias is the only interesting notion of security in this case.

As we said, we define the bias $\varepsilon$ as a multiplicative factor instead of additive, since the honest probabilities can be much different and in this case our definition makes more sense. To relate this bias to the one previously studied in coin flipping we have that coin flipping protocols with bias $\varepsilon \leq \sqrt{2} + \delta$ exist for any $\delta > 0$, see \cite{CK09}, and weak coin flipping protocols with bias $\varepsilon \leq 1 + \delta$ exist for any $\delta > 0$, see \cite{Moc07}.

\subsection{Extending Kitaev's Lower Bound to Forcing Oblivious Transfer}

We now extend Kitaev's formalism from the setting of coin flipping to the more general setting of  $\knfOT$. 

Suppose Alice and Bob have private spaces $\calA$ and $\calB$, respectively, and both have access to a message space $\calM$ each initialized in state $\ket{0}$. Then, we can define an $m$-round $\knfOT$ protocol using the following parameters:
\begin{itemize}
\item Alice's unitary operators $U_{A,1}, \ldots, U_{A,m}$ which act on $\calA \otimes \calM$
\item Bob's unitary operators $U_{B,1}, \ldots, U_{B,m}$ which act on $\calM \otimes \calB$
\item Alice's POVM $\set{\Pi_{A, abort}} \cup \set{\Pi_{A,x} : x \in \Z_2^n}$ acting on $\calA$, one for each outcome
\item Bob's POVM $\set{\Pi_{B, abort}} \cup \set{\Pi_{B,(b, x_b)} : b \text{ a k-element subset of n indices}, x_b \in \Z_2^k}$ acting on $\calB$, one for each outcome
\end{itemize}
We now show the criteria for which the parameters above yield a proper $\knfOT$ protocol. In a proper protocol we require that Alice and Bob's measurements are consistent and that the outcomes are uniformly random when the protocol is followed honestly.
Define 
\begin{equation*} 
\ket{\psi} := (I_{\calA} \otimes U_{B,m})(U_{A,m} \otimes I_{\calB}) \cdots (I_{\calA} \otimes U_{B,1})(U_{A,1} \otimes I_{\calB}) \ket{0}_{\calA \otimes \calM \otimes \calB} 
\end{equation*}
to be the state at the end of an honest run of the protocol. Then, we require the unitary and measurement operators to satisfy the following condition:
\begin{equation*} 
\norm{(\Pi_{A,x} \otimes I_{\calM} \otimes \Pi_{B,(b,x_b)}) \ket{\psi}}_2^2 = \dfrac{1}{\chews{n}{k} 2^n} \text{ for } (x,b,x_b) \text{ consistent. } 
\end{equation*}

Similar to coin flipping, we can capture cheating strategies as semidefinite programs. Bob can force Alice to output a specific ${x} \in \Z_2^n$ with maximum probability equal to the optimal value of the following semidefinite program
\begin{equation*} 
\begin{array}{rrrcllllllllllllll}
& B_{{x}} = \max                         & \inner{\Pi_{A,{x}} \otimes I_{\calM}}{\rho_{A,N}} \\
                     & \textrm{subject to} & \tr_{\calM}(\rho_{A,0}) & = & \kb{0}_{\calA} \\
                     &                               & \tr_{\calM}(\rho_{A,j}) & = & \tr_{\calM}(U_{A,j} \rho_{A,j-1} U^*_{A,j}), & \text{ for } j \in \set{1, \ldots, N} \\
                 &                                   & \rho_{A,0},  \ldots, \rho_{A,N} & \in & \Pos(\calA \otimes \calM), & \text{ for } j \in \set{0, \ldots, N}, 
\end{array} 
\end{equation*} 
where $\Pos(\calH)$ is the set of positive semidefinite matrices over the Hilbert space $\calH$. The states $\rho_i$ represent the part of the state under Alice's control after Bob sends his $i$'th message. The constraints above are necessary since Bob cannot apply a unitary on $\calA$. They are also sufficient since Bob can maintain a purification during the protocol consistent with the states above to achieve a cheating probability given by the corresponding objective value.

To capture Alice's cheating strategies we can do the same as for cheating Bob and examine the states under Bob's control during the course of the protocol. That is, Alice can force Bob to output a specific $k$-element subset ${b}$ and ${x_b} \in \Z_2^k$ with maximum probability equal to the optimal value of the following semidefinite program 
\begin{equation*} 
\begin{array}{rrrcllllllllllllll}
& A_{{b},{x_b}} = \max                         & \inner{I_{\calM} \otimes \Pi_{B,({b}, {x_b})} }{\rho_{B,N}} \\
                     & \textrm{subject to} & \tr_{\calM}(\rho_{B,0}) & = & \kb{0}_{\calB} \\
                     &                               & \tr_{\calM}(\rho_{B,j}) & = & \tr_{\calM}(U_{B,j} \rho_{B,j-1} U^*_{B,j}), & \text{ for } j \in \set{1, \ldots, N} \\
                 &                                   & \rho_{B,0},  \ldots, \rho_{B,N} & \in & \Pos(\calM \otimes \calB), & \text{ for } j \in \set{0, \ldots, N}. 
\end{array} 
\end{equation*} 

The proofs that these capture the optimal cheating probabilities are the same as those used for coin flipping in \cite{Kit03} and \cite{ABD+04}. Using these semidefinite programs we can prove the following theorem.

\vspace*{12pt}
\noindent
{\bf Theorem~3:}
In any $\knfOT$ protocol and consistent $b,x,x_b$ (that is, when $x_b$ is the $b$ entries of $x$) we have
\begin{equation*} 
B_{{x}} \cdot A_{{b}, {x_b}} \geq \Pr[\text{Alice honestly outputs } {x} \text{ and Bob honestly outputs } ({b}, {x_b})] = \dfrac{1}{\chews{n}{k} 2^n}. 
\end{equation*}
In particular, the forcing bias satisfies $\varepsilon \geq \sqrt{2}^k$. 
\vspace*{0pt}

Once we extended the semidefinite programming formulation, the proof of the theorem follows almost directly from the proof in \cite{Kit03} and \cite{ABD+04} for coin flipping except that the honest outcome probabilities are different in our case. Namely, for $\ket{\psi}$ defined above, we have
\begin{equation*} 
\norm{(\Pi_{A,x} \otimes I_{\calM} \otimes \Pi_{B,(b,x_b)}) \ket{\psi}}_2^2 
= \dfrac{1}{\chews{n}{k} 2^n} 
\end{equation*}
when $x$, $b$, and $x_b$ are consistent and $0$ otherwise.


\subsection{A Protocol with Optimal Forcing Bias}

In this section we prove Theorem~4. First, consider the following protocol which achieves the bound in Theorem~3 but is asymmetric. Alice sends $n$ random bits to Bob. Bob, then, outputs $b$, a random $k$-index subset of $n$ indices, and $x_b$.
In this protocol Bob can force a desired outcome with probability $\frac{1}{2^n}$ and Alice can force a desired outcome with probability $\frac{1}{\chews{n}{k}}$. Thus the product of the cheating probabilities is optimal, that is it achieves the lower bound in Theorem~3. However the protocol is asymmetric. This can be easily remedied using coin flipping. We present an optimal protocol with this security definition.

\vspace*{12pt}
{ \center
\begin{tabular}{|l|}
\hline \\
\quad \quad \underline{An Optimal $\knfOT$ Protocol with Forcing Bias $\sqrt{2}^k$} \\
\quad \\
\quad $1.$ Bob outputs a random index set $b$ of $k$ indices and sends the result to Alice. \qquad \\
\quad \\
\quad $2.$ Alice and Bob play a coin flipping game with bias $\sqrt{2} + \delta$ \\
\quad $\phantom{2.}$ (for a $\delta > 0$ sufficiently small) to determine each bit in $x_b$. \\
\quad \\
\quad $3.$ Alice randomly chooses her bits not in $b$. \\
\quad \\
\hline
\end{tabular}
\quad \\
}
\vspace*{12pt}

\vspace*{12pt}
\noindent
{\bf Theorem~4:}
For any $\gamma > 0$ we can choose a $\delta > 0$ such that the $\knfOT$ protocol above satisfies 
\begin{equation*} 
A_{b, x_b} \leq \dfrac{\sqrt{2}^k (1 + \gamma)}{\chews{n}{k} \cdot 2^k} \quad \mbox{and} \quad B_x  \leq \dfrac{\sqrt{2}^k (1 + \gamma)}{2^n}. 
\end{equation*}

\vspace*{0pt}
\noindent
{\bf Proof:}
Fix $\gamma > 0$ and a coin flipping parameter $\delta > 0$ such that $\left( \frac{1}{\sqrt{2}} + \frac{\delta}{2} \right)^k \leq \frac{\sqrt{2}^k(1 + \gamma)}{2^k}$. This can be achieved by taking $\delta = O(\frac{\gamma}{k})$. This sets an upper bound on the probability of forcing a $k$ bit-string using $k$ coin flipping protocols each with a maximum cheating probability of $\frac{1}{\sqrt{2}} + \frac{\delta}{2}$. We now analyze each party cheating. For Alice cheating, she has no control over the index set but she can try to force a particular bit-string for $x_b$. Her maximum cheating probability is 
\begin{equation*} 
\frac{1}{\chews{n}{k}} \cdot \left( \frac{1}{\sqrt{2}} + \frac{\delta}{2} \right)^k \leq \frac{1}{\chews{n}{k}} \cdot \frac{\sqrt{2}^k(1 + \gamma)}{2^k} = \frac{\sqrt{2}^k(1 + \gamma)}{\chews{n}{k} 2^k}. 
\end{equation*}
Bob has no control over Alice's $n-k$ remaining bits so Bob can cheat with maximum probability 
\begin{equation*} 
\frac{1}{2^{n-k}} \cdot \left( \frac{1}{\sqrt{2}} + \frac{\delta}{2} \right)^k \leq \frac{1}{2^{n-k}} \cdot \frac{\sqrt{2}^k(1 + \gamma)}{2^k} = \frac{\sqrt{2}^k(1 + \gamma)}{2^n}. \quad \square 
\end{equation*}

For the special case of $\fOT{1}{2}$ we have the following corollary.

\vspace*{12pt}
\noindent
{\bf Corollary~19:}
[Optimal $\fOT{1}{2}$]  
There exists a $\fOT{1}{2}$ protocol where each party has honest outcome probabilities of $1/4$ and neither party can cheat with probability greater than $\frac{1}{\sqrt{8}}(1 + \gamma)$, for any $\gamma > 0$. 
\vspace*{12pt}
 
Note that we have strong coin flipping protocols with $poly(m)$ rounds that achieve $\delta = \frac{1}{poly(m)}$. Hence, our protocol also achieves $\gamma = \frac{1}{poly(m)}$ with $poly(m)$ rounds.

Last, we remark that this protocol is completely classical with the exception of the quantum coin flipping subroutines. This is similar to the optimal coin flipping protocol in \cite{CK09} designed using classical messages and optimal quantum weak coin flipping subroutines.

\section*{Acknowledgements}
\noindent
IK was partly supported by the Agence Nationale de la Recherche under
the project ANR-09-JCJC-0067-01.
JS acknowledges support from NSERC, MITACS, ERA (Ontario), and the EU-Canada Exchange Program.

\noindent

\nocite{SDP}
\nocite{C07}
\nocite{BCS12}
\bibliographystyle{alpha}
\bibliography{paper}

\end{document}